\documentclass[pra,aps,showpacs,superscriptaddress,twocolumn,floatfix]{revtex4}
\usepackage{graphicx}
\usepackage[ansinew]{inputenc}
\usepackage{array}
\usepackage{color}
\usepackage{amsmath}
\usepackage{amsxtra}
\usepackage{amstext}
\usepackage{amssymb}
\usepackage{latexsym}

\begin{document}

\title{Quantum measurement backaction from a BEC coupled to a mechanical oscillator} 
\author{S. K. Steinke}
\author{S. Singh}
\author{M. E. Tasgin}
\author{P. Meystre}
\affiliation{B2 Institute, Department of Physics and College of Optical Sciences\\The University of Arizona, Tucson, Arizona, 85721.}

\author{K. C. Schwab}
\affiliation{Applied Physics, California Institute of Technology, MC 128-95, Pasadena, California 91125.}

\author{ M. Vengalattore}
\affiliation{Laboratory of Atomic and Solid State Physics, Cornell University, Ithaca, New York, 14853.}

\date{\today}

\begin{abstract}
We study theoretically the dynamics of a a hybrid optomechanical system consisting of a macroscopic mechanical membrane magnetically coupled to a spinor Bose-Einstein condensate via a nanomagnet attached at the membrane center. We demonstrate that this coupling permits us to monitor indirectly the center-of-mass position of the membrane via measurements of the spin of the condensed atoms. These measurements normally induce a significant backaction on the membrane motion, which we quantify for the cases of thermal and coherent initial states of the membrane. We discuss the possibility of measuring that quantum backaction via repeated measurements. We also investigate the potential to generate non-classical states of the membrane, in particular Schr{\"o}dinger cat states,  via such repeated measurements.
\end{abstract}


\maketitle

\section{Introduction}
While the foundations of Quantum Measurement Theory were already laid down in the early days of quantum mechanics, quantitative studies are a relatively recent development~\cite{Braginsky, KippenbergScience09}. These studies are now driving significant advances in quantum information science and quantum metrology, with major progress initiated by the desire to test Bell's inequalities on the one hand, and by the quest for gravitational wave detection on the other~\cite{Aspect82,Hollenhorst}.  A common aspect of these and related studies is the need to quantify, control, and possibly exploit the quantum backaction of one or a series of measurements on a quantum mechanical system. 

An important development in this context is the rapid progress witnessed by cavity optomechanics, which makes it increasingly realistic to consider the use of mechanical systems operating in the quantum regime to make precise and accurate measurements of feeble forces and fields \cite{KippenbergScience09}. In many cases, these measurements amount to the detection of exceedingly small displacements. In this context, {\it hybrid systems} consisting of coupled atomic (or molecular) and nanomechanical systems may prove particularly useful. The robust and scalable infrastructure provided by NEMS/MEMS devices, coupled with the high precision measurement capability of quantum gases~\cite{Vengalatorre2007, Obrecht2007, Zoest2010}, makes them an attractive combination for sensitive force measurements, as well as for a quantitative study of dissipation and decoherence processes at the quantum-classical interface.  As a result, there are ongoing experimental~\cite{Kitching2006, Treutlein2010} and theoretical~\cite{Treutlein2007, Genes2008, Hammerer2009, SinghPM2010, Hammerer2010} efforts toward coupling mechanical systems to atomic ensembles.

\begin{figure}[ht]
\begin{center}
\includegraphics[width=2.7in]{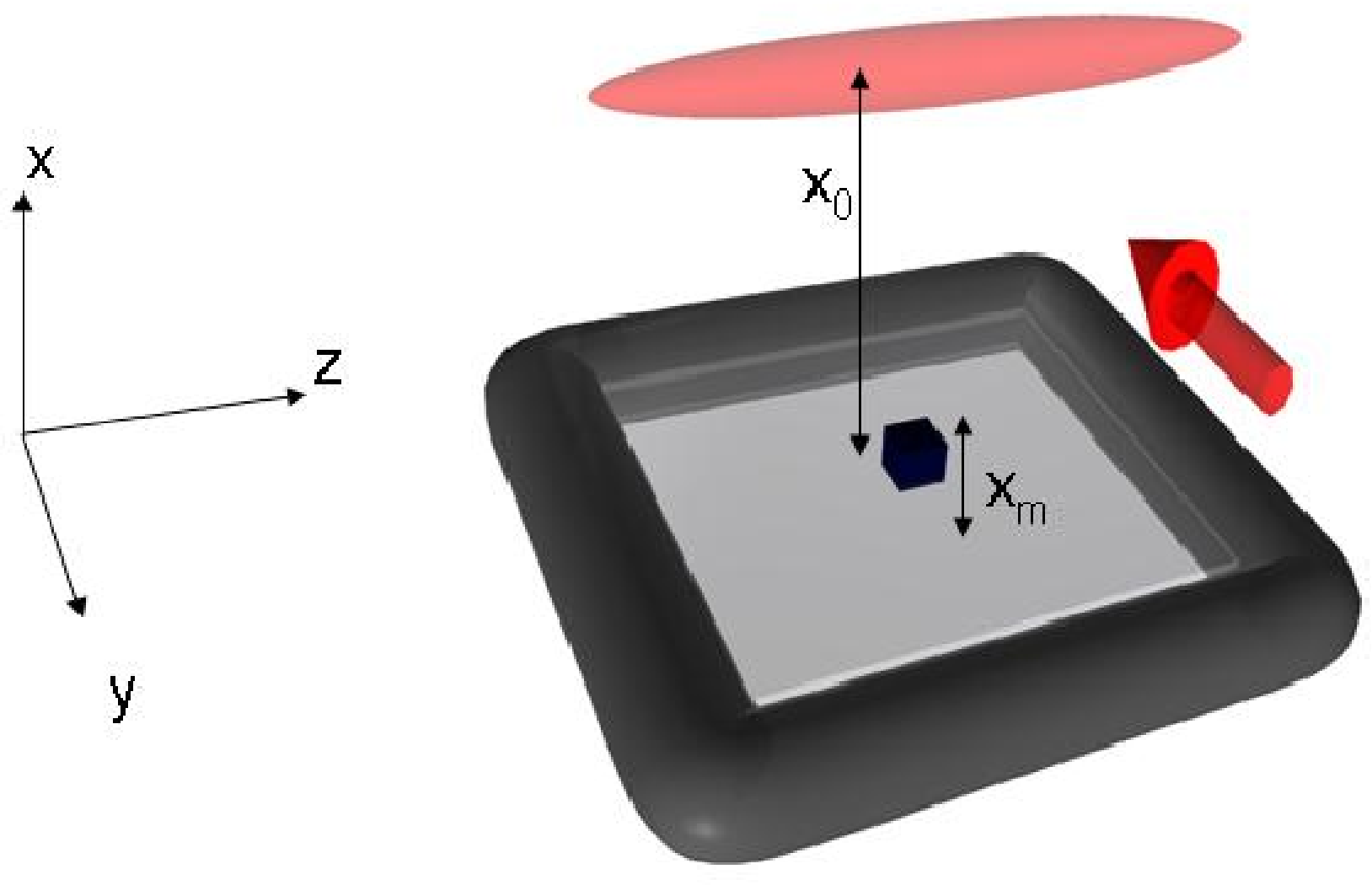}
\end{center}
\caption{Proposed experimental setup for backaction detection, involving a magnetic particle located at the center of a vibrating membrane and polarized along the $z$-axis. This setup produces a spatially inhomogeneous magnetic field that is detected by the BEC. The BEC's long axis is along the $z$-direction, the probe light (red arrow) is along $y$, and the membrane oscillations are along the $x$-axis.}
\label{fig:Setup}
\end{figure}

The system that we consider in this paper consists of a mechanical membrane magnetically coupled to a spinor Bose-Einstein condensate, an arrangement somewhat similar to a system previously considered in Refs.~\cite{Treutlein2007,SinghPM2010}. The magnetic coupling, generated via a nano-magnet anchored on the membrane, entangles the membrane to the spin state of the BEC. We may then make an indirect observation of the center-of-mass position of the membrane through a measurement of the spins of the condensed atoms. While such magnetic coupling and concomitant entanglement is possible between the membrane and a thermal gas of spinor atoms (or even a single atom), the BEC has two advantages over the alternatives when used as a position sensor. Namely, bosonic stimulation between the atoms leads to increased signal strenth, and two-body collisions are suppressed, giving rise to a much longer phase coherence time when compared to thermalized atoms. Now, the indirect position measurement can be achieved generally speaking in two ways: through a strong, or projective, measurement of the spin state of the BEC, or through a weak, or dispersive, measurement of the spins. An example of the former would be through a Stern-Gerlach-type measurement. Here, the atoms are subjected to a spatially inhomogenous field seperating out the different spin components. The different spin populations can then be measured via a standard absorptive imaging technique. A likely candidate for the latter type of measurement would be observation of the Larmor frequency through a phase contrast imaging technique, as described in Ref.~\cite{Vengalatorre2007}. Since the Larmor frequency is proportional to the local magnetic field -- which is modulated by the motion of the magnetic domain attached to the oscillating membrane -- it provides a measure of the membrane motion.  Either type of measurement of the BEC spin can induce a backaction on the membrane, modifying its position and/or momentum in proportion to the strength of the measurement. The main goal of this project is thus to quantify the effect of using the BEC as a position sensor on the membrane and to evaluate the feasibility of measuring the backaction of such a measurement. In the case where the backaction is small we can indeed use the BEC to make ultraprecise measurements of the membrane, and when it is large, it may provide a means to observe the effects of a quantum measurement on a macroscopic object. So far such a quantum effect has only been observed in dilute and isolated systems such as ultracold gases~\cite{Murch2008}.

Due to the scope of this endeavor, we find it convenient to divide our results into three papers. In this, the first paper, we present the experimental setup and develop a simplified Hamiltonian to describe the interaction of the membrane and condensate. We derive a general formulation of the system state after a single or multiple arbitrary measurements of the condensate's spin state, and then present results for specific cases where the measurements are projective. We show that in principle the membrane can be put into a non-classical state by such a procedure. In the second paper we extend these considerations to weak measurements, focusing on quantifying the backaction of indirect observation of the Larmor frequency via phase contrast imaging. In the third paper we will exploit the backaction of repeated measurements and coherent control to achieve quantum control of the state of the membrane, allowing in principle the production of squeezed or Schr{\"o}dinger cat states. Because the effects of environmental coupling, i.e. decoherence and dissipation, are critical to coherent control protocols, we will also include these effects in the computations of paper 3.

This paper is organized as follows; Section II introduces the system under consideration and derives a model Hamiltonian that describes the magnetic coupling of the membrane to the condensate. Section III  discusses the indirect measurement of the membrane center-of-mass position via a strong measurement of the spin of the condensate, assuming that the membrane is initially in a thermal state.  In particular, we present the post-measurement Wigner function of the membrane and interpret its main features in terms of a backaction parameter. Section IV then turns to the case of repeated measurements on this system, illustrating how a sequence of measurements provides a direct signature of the backaction. Section V discusses the possibility of producing highly non-classical states of the membrane, considering specifically the case where it is initially in a coherent state. We show that repeated measurements typically leave the membrane in a non-classical state characterized by a non-positive Wigner distribution. Finally, Section VI is a summary and outlook. Some technical details of the calculations and experiment are included as appendices.

\section{Model}
As mentioned in the introduction, the system under consideration consists of a micromechanical membrane of fundamental mode frequency $\omega_m$ and effective mass $m$, whose center-of-mass is oscillating, perhaps under the influence of a weak force. Our goal is to develop and analyze a quantum measurement scheme that permits us to characterize the center-of-mass motion of that membrane.

The scheme that we envisage involves integrating this membrane into a hybrid system whose other component is an elongated spinor Bose-Einstein condensate with long axis $z$, see Fig.~1. The condensate is subjected to a static magnetic field $B_0$ along the quantization axis $z$. The membrane is magnetically coupled to the condensate via a magnetic domain anchored at the center of the membrane.

We assume for simplicity that the magnetic domain is a point dipole located at the origin and polarized along the $z$-axis, ${\bf \mu_m} = \mu_m{\bf\hat{z}}$ . The magnetic field due to that dipole is
\begin{equation}
 {\bf B(r)}=\frac{\mu_0}{4 \pi}\frac{1}{r^3}\left(3({\bf{\mu_m}\cdot\bf\hat{r}}){\bf\hat{r}}-{\bf \mu_m}\right).
\label{B}
\end{equation}
The small spatial variation of the resulting magnetic field along the long axis of the condensate results in a variation of its Larmor precession frequency. As we show in the following, this dependence permits us to characterize the expectation value of center-of-mass mode of oscillation of the membrane. 

The Zeeman interaction between a single atom in the condensate and the total magnetic field ${\bf B}_0 + {\bf B}({\bf r})$ is described by the interaction Hamiltonian
\begin{eqnarray}
V&=&-{\bf \mu_a} \cdot ({\bf B}_0 + {\bf B}({\bf r}) )\nonumber \\
\label{Hmag}
&=&\mu_B g_F\left( F_x B_x + F_yB_y+F_z(B_0+B_z)\right),
\end{eqnarray}
where $F_i$ is the $i^{th}$ component of the spin-1 operator, $\mu_B$ is the Bohr magneton, and $g_F=2$ for alkali atoms. Because of the extremely low temperature of the condensate and the relatively small trapping frequencies employed (order 10 -- 100 Hz), we will neglect the kinetic energy of the atoms and the trapping potential in our model. As a result of the spatial dependence of the magnetic field, atoms at different positions along the long axis of the condensate precess at different frequencies and dephase over a period of time. It is this $z$-dependent phase difference that is picked up in spin population measurements.

We note that the transverse dependence of the magnetic field also results in an inhomogeneous broadening of the Larmor frequency. The effects of this broadening can be minimized, the more elongated along $z$ and tightly confined in the transverse directions the condensate is. In the following we consider for simplicity a condensate that is almost one-dimensional and confined to a region close to $y=0$, so that $B_y\approx 0$ and $F_y B_y \approx 0$. Furthermore, close to $z=0$, i.e. for the fraction of the BEC directly above the dipole, the magnetic field ${\bf B}({\bf r})$ is predominantly in the $z$- direction provided that $x_0$, the equilibrium distance between the BEC and the membrane, is much greater than the relevant coordinates $y$ and $z$. In that case we can ignore the effects of $B_x$ and $B_y$ altogether. As shown in detail in Appendix A, for small displacements $x_m$ of the membrane compared to $x_0$, the single atom magnetic coupling Hamiltonian (\ref{Hmag}) reduces then to
\begin{equation}
V=\mu_B g_{F}F_z\left[B_0+\frac{\mu_0\mu_m}{4\pi x_0^{4}}\left(-x_0+ 3x_m\right)\right].
\end{equation}
The first and second terms of this expression are independent of time, while the third term, proportional to $x_m$, varies sinusoidally in time. We exploit this property by rewriting the magnetic Hamiltonian as
\begin{equation}
\label{eq:simpleAHamiltonian}
V=\mu_B g_F F_z \left( B_c+B_v' x_m \right),
\end{equation}
where 
\begin{eqnarray}
B_c&=&B_0-\frac{\mu_o \mu_m}{4\pi x_0^3},\nonumber\\
B_v'&=&\frac{3 \mu_o \mu_m}{4\pi x_0^4}.
\end{eqnarray}

By treating the condensate as $N$ non-interacting spin-1 atoms under the mean-field approximation and by assuming that the magnetic field acting on the atoms in the detector region is roughly constant, we can easily arrive at the many-atom magnetic interaction Hamiltonian. It is simply
\begin{eqnarray}
\label{eq:V}
V_\mathrm{BEC} &=& \int d^3\vec x\Psi^\dag(\vec x)V\Psi(\vec x)\nonumber\\
 &=&N\mu_Bg_F F_z (B_c +B_v'x_m),
\end{eqnarray}
where $\Psi(\vec x)$ is the annihilation operator for an atom at position $\vec x$ (i.e. it is a Schr\"{o}dinger field operator). The total system Hamiltonian of the hybrid BEC-membrane system is then
\begin{equation}
 H= H_m+ V_\mathrm{BEC}.
\end{equation}
Here
$$
H_m=p^2/2m+m\omega_m^2x^2/2
$$
is the membrane Hamiltonian, and we have dropped (and will continue to drop) the subscript in membrane displacement for compactness, $x_m\rightarrow x$. By explicitly combining both terms of $H$ and completing the square for $x$, we express the total system Hamlitonian as
\begin{equation}
\label{eq:Hamiltonian}
H=\frac{p^2}{2m}+ \frac{1}{2}m\omega_m^2\left(x+A F_z\right)^2+
\left\{\hbar N\Omega_{L0} F_z-\hbar\delta \Omega F^2_z\right\}
\end{equation}
where
\begin{equation}
\label{eq:baction}
A=\mu_B g_F N B_v'/m\omega^2,
\end{equation}
which we call the {\em backaction parameter} in anticipation of the following sections,
\begin{equation}
\Omega_{L0} = \mu_B g_F B_c/\hbar,
\end{equation}
which is simply the Larmor precession frequency of the atomic spins, and 
\begin{equation}
\delta\Omega=m\omega_m^2A^2/2\hbar.
\end{equation}

\section{Spin measurement}

In this and the following sections, we evaluate the backaction on the membrane of an arbitrary measurement operator that acts on the spin degree of freedom of the BEC. We consider the cases of a single and multiple spin measurements for a membrane initially prepared either in a thermal or in a coherent state. The analytical expressions provided for the post measurement density matrix (and subsequently the Wigner function) are for a general measurement operator. However, to illustrate the backaction effects in an intuitive manner, we consider in this paper the particular case of simple projective measurements of $F_y$, corresponding to all atoms being in the $F_y=0,+1$ or $-1$ state. While this is not the most likely outcome of a typical Stern-Gerlach measurement, it does serve as a simple demonstration of backaction induced dynamics.

The measurement proceeds in the following way: at time $t=0$ a $\pi/2$ pulse is applied to the condensate, preparing all atoms in $|F_x=1\rangle$ state. Following that preparation stage, the atomic spins precess about the $z$-axis until a later time $t_1$ when the spin population measurement is made. Additional measurements can be performed at later time intervals $t_i, i=2, 3, \ldots$. Further precise details of the experimental setup are presented in Appendix D.

Equation~(\ref{eq:Hamiltonian}) already gives a clear indication of the backaction of the measuring apparatus -- the condensate --  on the membrane. As a result of their coupling the membrane Hamiltonian is modified from being a harmonic oscillator centered at the origin to one that is shifted by the quantity $A F_z$, indicating that the backaction depends on the outcome of a specific spin measurement. Here, we give an explicit description of the measurement process by evaluating the pre-- and post-measurement density operator of the membrane, and the corresponding Wigner function.

\subsection{Density operator}

We assume that the membrane and the condensate are initially uncorrelated, 
$$
\rho=\rho_m(0)\otimes \rho_{\rm BEC}(0)
$$ 
and denote the initial density matrix elements of the BEC as $ \rho_{\alpha,\beta}^0$, where $\rho_{\alpha,\beta}=\langle \alpha|\rho_{\rm BEC}|\beta \rangle$, and  $\alpha,\beta$ are the various spin states, $\alpha,\beta = \{0,\pm1\}$.  We also assume for now that the membrane center-of-mass is initially in thermal equilibrium with temperature $T$. Defining
$$
\eta =\hbar \omega_m/2 k_B T,
$$
its density matrix elements in position space are then
\begin{widetext}
\begin{equation}
\label{eq:thermalHO}
\rho_m(x_f,x_i,t=0)=\langle x_f |\rho_m |x_i \rangle=\sqrt{\frac{m\omega_{m}}{\pi \hbar}\tanh{\eta}} 
\times \exp\left[-\frac{m\omega_{m}}{4 \hbar}\left((x_f+x_i)^2\tanh{\eta} + (x_f-x_i)^2\coth{\eta}\right)\right].
\end{equation}

For $t>0$ the spin components of the condensate undergo a Larmor precession about the $z$-axis. Since  $[H_{\rm BEC},H_m+V]=0$, we can use the Baker-Hausdorff relation to re-express the propagator $U(t) = \exp(-i Ht/\hbar)$ as
$$
U(t)=e^{-it H_{\rm BEC}/\hbar} e^{-it(H_{m}+V)/\hbar}.
$$
This allows us to find the evolution of the system density matrix in a straightforward way. After an interaction time $t_1$, this evolution results in the matrix elements of the density operator of the membrane + condensate system to become
\begin{eqnarray}
\label{IntSysDMatrix}
&& \langle\alpha,x_f|\rho(t)|\beta,x_i\rangle= \rho^0_{\alpha \beta} \exp[-i\Omega_{L0}(\alpha-\beta)t_1 +i\delta\Omega_{L}(\alpha^2-\beta^2)t_1]\sqrt{\frac{m\omega_m}{\pi \hbar}\tanh{\eta}}\times\\ \nonumber
&& \exp\left[-\frac{m \omega_m}{4\hbar} \left\{ \left(x_f+x_i+(\alpha+\beta)A(1-\cos{\omega_m t_1})\right)^2\tanh{\eta}+\left(x_f-x_i+(\alpha-\beta) A(1-\cos{\omega_m t_1})\right)^2\coth{\eta}\right.\right.\\ \nonumber
&& \left. \left.+4iA\sin{\omega_m t_1}(\alpha x_f-\beta x_i)+2iA^2(\alpha^2-\beta^2)\sin{\omega_m t_1}(2-\cos{\omega_m t_1})\right\}\right].
\end{eqnarray}
\end{widetext}
That is, the interaction of the membrane with the BEC displaces its center-of-mass motion in both position and momentum by amounts that depend explicitly on the spin components $\alpha$ and $\beta$, as well as on the backaction parameter $A$.

\subsection{Single measurement}

As already discussed a measurement of arbitrary type, but dependent on $F_y$, is carried out on the BEC at time $t_1$. The post-measurement density matrix of the membrane depends on the measurement outcome, and is given by
\begin{equation}
\label{eq:condHOdensity}
\langle x_f|\rho_{m}|x_i \rangle_\phi=\frac{1}{P(\phi)} \mathrm{tr_{_{BEC}}}(W^\phi \rho_\mathrm{sys}).
\end{equation}
Here $\phi$ is the outcome of the $F_y$ measurement, $P(\phi)$ is the probability of that outcome, $W^\phi$ is the Kraus operator corresponding to the effects of the measurement on the BEC's quantum state, and $\rho_\mathrm{sys}$ is the complete system density matrix given in~\eqref{IntSysDMatrix}. Though Eq.~\eqref{eq:condHOdensity} is true for arbitrary measurements (i.e. for arbitrary Kraus representations), in later plots and numerics, we take the operators $W^\phi$ to be projectors onto the eigenstates of $F_y$, namely, the 3 operators $|F_y=\gamma\rangle\langle F_y=\gamma|$, $\gamma=\{0,\pm1\}$. This simplified situation displays the salient features of measurement back-action without requiring extensive computations. Such a scenario could be realized through e.g. a Stern-Gerlach aparatus. As mentioned above, we will return to the precise computation of operators $W^\phi$ corresponding to the planned experimental setup (Appendix D) in subsequent work.

\subsection{Phase Space Representation}

The effect of the measurement on the state of the membrane can be visualized particularly clearly in terms of its Wigner distribution function
\begin{equation}
\label{eq:WignerDef}
W(x,p)=\frac{1}{2 \pi \hbar}\int d\xi e^{-ip\xi/ \hbar}\langle x+\xi/2| \rho | x-\xi/2\rangle
\end{equation}
For a harmonic oscillator in a thermal state, we have
\begin{widetext}
\begin{equation}
W(x,p,t=0)=\frac{1}{\pi\hbar}\tanh{\eta}\exp\left[-\frac{m \omega_m}{\hbar} \left(x^2+\left(\frac{p}{m\omega_m}\right)^2\right)\tanh{\eta}\right]
\end{equation}
an expression that should be contrasted to the post-measurement Wigner function, which is found to be
\begin{eqnarray}
\label{eq:WignerPM}
W(x,p,t_1)&=&\frac{1}{\pi\hbar}\tanh{\eta} \frac{1}{P(\phi)} \sum_{\alpha,\beta}W^\phi_{\beta\alpha}\rho^0_{\alpha \beta} \exp{[-i\Omega_{L0}(\alpha-\beta)t_1 +i \delta \Omega_L(\alpha^2-\beta^2)t_1]} \times \\ \nonumber
&&\exp\left[-\frac{m \omega_m}{\hbar}\left\{\left(\left(x+\frac{A}{2}(\alpha+\beta)(1-\cos(\omega_m t_1))\right)^2+\left(\frac{p}{m\omega_m}+\frac{A}{2}(\alpha+\beta)\sin(\omega_m t_1) \right)^2\right)\tanh{\eta} \right.\right.\\ \nonumber
&& \left.\left.+iA(\alpha-\beta)\left(x\sin(\omega_m t_1)-\frac{p}{m\omega_m}\right (1-\cos(\omega_m t_1)\left )+\frac{A}{2}(\alpha+\beta)\sin(\omega_m t_1)\right)\right\} \right].
\end{eqnarray}
\end{widetext}
At time $t_1$ the first line of Eq.~(\ref{eq:WignerPM}) simply describes the imposition, due to the evolution of the BEC, of a phase on each term of $W(x,p)$ which is dependent on the measurement outcome and spin indices. More interesting are the last two lines in that expression: The second line describes a term-by-term shift in the initial Gaussian probability distribution The new phase-space center of each term in the Wigner function depends on the spin indices of that term, and it rotates through phase space at the membrane frequency. 

As the third line in $W(x,p,t_1)$ is imaginary, it results in interference between the terms of the Wigner function (visible below in Fig.~\ref{fig:measureFy1diffA}). From inspection of Eq.~\eqref{eq:WignerPM}, we see that the interference will increase with larger $A$.  The physical significance of this interference can be understood from the fact that the initial state of the BEC, $|F_x=1\rangle$, is not an eigenstate of the interaction Hamiltonian (proportional to $F_z$). So, one can think of the condensate as experiencing three interaction Hamiltonians simultaneously, one for each of its spin components, and it is the interference between them that leads to the oscillations. We remark that the oscillations in the Wigner function can be seen as long as 
\begin{equation}
\label{eq:Wosc}
\frac{A}{x_{zp}}\stackrel{>}{\sim} \sqrt{\tanh{\eta}}
\end{equation}
where $x_{zp}=\sqrt{\hbar/2m\omega_m}$.

\subsection{Backaction}

\begin{widetext}
\begin{figure}[t!!]
\begin{center}
$\begin{array}{ccc}
\includegraphics[width=2.1in]{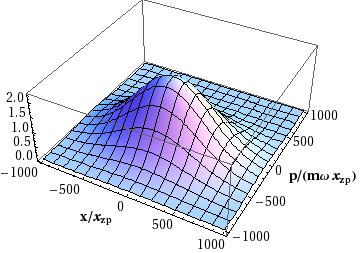}  &
\includegraphics[width=2.1in]{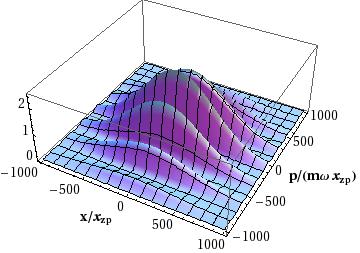} &
\includegraphics[width=2.1in]{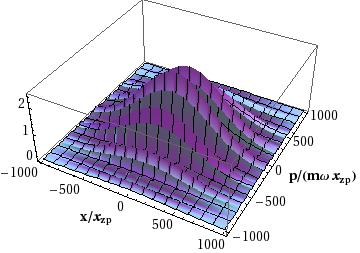} \\[0.0cm]
\mbox{\bf a. $A=0.01A_0$} &\mbox{\bf b. $A=0.1A_0$} &\mbox{\bf c.  $A=A_0$}
\end{array}$
\end{center}
\caption{(Color online) Wigner distribution function of the membrane after one measurement giving the result $F_y=1$, for several values of the backaction parameter: (a) $A=0.01 A_0$: (b) $A=0.1A_0$ and (c) $A=A_0$, with $A_0=0.22 \sqrt{\hbar/2 m\omega_m}$. The evolution time is $t=\pi/\omega_m$ in all three cases. The Wigner function has been multiplied by $10^6$ to make the axes legible.}
\label{fig:measureFy1diffA}
\end{figure}
\end{widetext}

To illustrate the effect of backaction, we assume the following membrane parameters for the remainder of the paper (unless stated otherwise): $\omega_m= 2\pi 10^6$ rad/s, $m=5\times10^{-13}$ kg, $\mu_m= 2\times10^{-11} {\rm A\cdot m^2}$, and an initial temperature of 4K. The static external magnetic field is $B_0=0.1$ Gauss, and the condensate is $x_0=5 \times 10^{-6}$ m away from the membrane, resulting in a single-atom backaction parameter $A_{\rm sa} = 9\times 10^{-21} $ m. We assume that $N=10^5$ atoms experience the same magnetic field in the detection region, yielding then an effective backaction parameter of $A_0=N A_{\rm sa}=9\times10^{-16}$m (for comparison, $A_0=0.22 x_{zp}$).  Fig.~\ref{fig:measureFy1diffA} shows the resulting post-measurement Wigner function for this specific value of $A_0$ (figure c), and for backaction parameters of $0.1 A_0$ (figure b), and $0.01 A_0$ (figure a), assuming that a measurement of $F_y$ with result 1 was made after an interaction time of $t_1=\pi
 /\omega_m$. 

As alluded to by Eq.~(\ref{eq:Wosc}), in order to observe the oscillations in the post-measurement Wigner function, we can either increase the temperature or increase $A$. However, increasing the temperature leads to dissipation and decoherence losses that are ignored in the present analysis, but result of course in a fast thermalization and associated smoothing of $W(x,p)$. As mentioned above, these effects will be discussed in future work. A more promising approach to observe quantum interference effects is to increase $A$ (Eq.~(\ref{eq:baction})), either by increasing the number of atoms in the effective detection zone, or by increasing $B_v'$ via a decrease of $x_0$. Since $B_v'$ scales as $1/x_0^4$, this may be the easiest way to reach the regime of observable Wigner function oscillations. Note however that for decreasing $x_0$ the simplified interaction Hamiltonian~(\ref{eq:V}) becomes less accurate as the components of the magnetic field along $x$ and $y$ become more important, so the approximations made in Eq.~\eqref{eq:Hamiltonian} will not be as valid.

The expectation value of the center-of-mass position $\langle x \rangle$ of a membrane in thermal equilibrium is zero, and its variance is
\begin{equation}
\sigma^2(x)=\langle x^2\rangle-\langle x \rangle^2=\langle x^2\rangle= \frac{\hbar}{2m\omega_m}\coth{\eta}.
\end{equation}
Immediately following the measurement, the membrane is no longer in thermal equilibrium, and $\langle x\rangle \neq 0$ in general. For large backaction parameters the oscillations in $W(x,p)$ become quite significant. In that regime the BEC is a poor position sensor, since its coupling to the membrane significantly perturbs the outcome of subsequent measurements, see Fig~\ref{fig:measureFy1diffA}. Here the measurement  creates a significant change in the phase-space distribution of the membrane that invalidates any
information gained about the position. However, as we show in Section V, a high value of $A$ is beneficial for preparing the membrane in highly non-classical states via repeated measurements. As with all schemes for state preparation involving repeated measurements, its not very efficient for highly excited initial states, as is the case in Fig.~\ref{fig:measureFy1diffA}  -- At 4 K, and for the membrane parameters of this example, the mean phonon occupation number is ~$8.3\times10^4$.

To investigate the efficiency of our setup as a position sensor, it is therefore appropriate to consider the limit of 
small $A_0$. Consider for concreteness the specific example where the outcome of the spin measurement is $F_y=1$. Ignoring then terms of order $A^2$, and for $t_1=\pi/\Omega_{L0}$, we find

\begin{equation}
\langle x\rangle=-\frac{A}{2}\sin({\omega_m \pi/\Omega_{L0}})\coth{\eta},
\end{equation}
with $\langle x^2\rangle$ remaining constant to lowest order in $A$. We then have 
\begin{equation}
\sigma^2(x)_{\rm pm}=\frac{\hbar}{2m\omega_m}\coth{\eta}-\frac{A^2}{4}\sin^2{(\omega_m \pi/\Omega_{L0})}\coth^2{\eta}
\end{equation}
where the subscript ``pm'' indicates post-measurement. The minimum backaction occurs for $\eta \rightarrow \infty$, or $T\rightarrow 0$. It also vanishes when the membrane frequency is an integer multiple of Larmor precession frequency,
\begin{equation}
\omega_m = n \Omega_{L0},
\end{equation} 
where $n$ in an integer, in which case it is possible to carry out stroboscopic QND measurements of the membrane position.

\section{Successive Measurements}

Generally, successive measurements of the condensate's spin result in an accumulation of backaction 
effects in the membrane. At the same time, they offer the potential for the coherent control of the center-of-mass motion of the membrane. To address such situations, we now consider the effect of a succession of measurements on the state of the membrane. The similarity between Eqs.~(\ref{eq:thermalHO}) and (\ref{IntSysDMatrix}) suggests that it should be possible to find a closed form for the density matrix of the composite system after an arbitrary number of measurements on the BEC. Indeed, this is the case and such a form is presented below.

\subsection{Analytical results}

We first consider the situation where the membrane begins in a thermal state, and assume that the interaction between the membrane and the BEC, (initially prepared in the $F_x=1$ state, is turned on at time $t_0=0$. After time $t_1$, a first measurement is performed on the spin of the BEC. The probe-BEC interaction time $t_{\rm meas}$ required to carry out that measurement is taken to be negligible compared to the other characteristic times of the system -- in practice this is true for phase contrast imaging.  The direct effect of the measurement is only on the BEC, and can be formally described by a Krauss operator $M^{(1)}$ that depends explicitly on the outcome of the measurement. Following that first measurement the system evolves unitarily for an additional time $t_2$, becoming re-entangled. A second measurement is then performed, acting on the BEC with an operator $M^{(2)}$, and this process is repeated $n$ times. The set of operators $M^\phi$ are related to the above $W^\phi$ by $W^\phi=M^{\phi\dag}M^\phi$.

Immediately following the $n$th measurement, the elements of the BEC-membrane system's density matrix are given by (see Appendix B for more details)
\begin{widetext}
\begin{eqnarray}
\label{eq:FinalDMatrix}
\langle \alpha,x_f |\rho(t)|\beta,x_i\rangle &=&\sum_{\alpha_1,...\alpha_n;\beta_1,...\beta_n}M^{(n)}_{\alpha\ \alpha_n}
M^{(n-1)}_{\alpha_n\alpha_{n-1}}...M^{(1)}_{\alpha_2\alpha_1}\rho_{\alpha_1\beta_1}
M^{(1)\dag}_{\beta_1\beta_2}...M^{(n-1)\dag}_{\beta_{n-1} \beta_n}
M^{(n)\dag}_{\beta_n \beta}\nonumber\\
&\times&\exp\left[-i\sum_{i=1}^n t_i\{\Omega_{L0}(\alpha_i-\beta_i)-\delta\Omega_{L}(\alpha_i^2-\beta_i^2)\}\right ]\sqrt{\frac{m\omega_m}{\pi\hbar}\tanh\eta}\nonumber\\
&\times&\exp\left[-\frac{m\omega_m}{4\hbar}\left\{(x_f+X[\alpha]A+x_i+X[\beta]A)^2\tanh\eta+(x_f+X[\alpha]A-x_i-X[\beta]A)^2\coth\eta\right. \right. \nonumber\\
&+&\left. \left. 4iA(P[\alpha](x_f+X[\alpha]A)-P[\beta](x_i+X[\beta]A))+2iA^2(\phi[\alpha]-\phi[\beta])\right\}\rule{0pt}{12pt}\right]
\end{eqnarray}
\end{widetext}
where
\begin{eqnarray}
\label{eq:DefineM}
M^{(i)}_{\sigma \tau}&=&\langle\sigma|M^{(i)}|\tau\rangle,\\
\label{eq:DefineX}X[\sigma] &=&\sigma_n-\sum_{i=1}^n\left(\sigma_i-\sigma_{i-1}\right)\cos T_{i,n},\\
\label{eq:Defineq}P[\sigma] &=&\sum_{i=1}^n\left(\sigma_i-\sigma_{i-1}\right)\sin T_{i,n},\\
\label{eq:DefinePhi}\phi[\sigma]&=&\sum_{i=1}^n(\sigma_i- \sigma_{i-1})^2 \sin T_{i,n} \cos T_{i,n}\nonumber\\
&\hspace{-42pt}+&\hspace{-26pt}2\sum_{i=1}^{n-1}\hspace{-8pt}\sum_{\ \ \ j=i+1}^n\hspace{-8 pt}(\sigma_i\hspace{-2pt}-\hspace{-2pt}\sigma_{i-1})(\sigma_j\hspace{-2pt}-\hspace{-2pt}\sigma_{j-1})\sin T_{j,n}\cos T_{i,n},\\
\label{eq:DefineTij}
T_{i,j}&=&\omega_m\sum_{k=i}^j t_k \mathrm{\ \ if\ } i\le j \mathrm{\ and\ } = 0 \mathrm {\ \ otherwise}.
\end{eqnarray}

The sums on $\alpha_i$ and $\beta_i$ run over $\{-1, 0, +1\}$, and in Eqs.~\eqref{eq:DefineX} -- \eqref{eq:DefinePhi} we have $\alpha_0=\beta_0=0$ wherever they appear. To be clear, in the equations above, the quantities $t_i$ are the intervals between measurements rather than the times themselves. Also, care should be taken as $X$, $P$, and $\phi$ are functions of the spin indices and thus must be recomputed for each term in the sum. Note also that the density matrix~(\ref{eq:FinalDMatrix}) is unnormalized. Its trace is equal to the probability of the particular sequence of measurement outcomes described by the specific set of operators $\{ M^{(i)}\}$ occurring in it.

A few remarks are in order before turning to a discussion of numerical results: First, the density matrix of the membrane can be obtained from Eq.~(\ref{eq:FinalDMatrix}) by a partial trace. It is the sum of many different contributions from various shifted (in position and momentum) thermal ensembles. A (numerical) computation problem does arise, though, as calculating the density matrix for $n$ measurements requires summing over $9^n$ terms. In our numerics we are thus restricted to few-measurement scenarios. The timing of the measurements is also very important to the BEC functioning as a detector. In one extreme case, if the measurements are made exactly at the natural frequency of the membrane, the interaction will be completely masked, except for the tiny second-order effect of $\delta\Omega_{L}$. This supports the claim that position is very nearly a stroboscopic QND variable.

The structures of Eqs.~(\ref{eq:FinalDMatrix}) and (\ref{eq:DefineX})-(\ref{eq:DefinePhi}) indicate that the results of the earliest measurements continue to be as important as those of later measurements. This is because in the absence of dissipation, there is no attenuation of the information gained nor of the backaction induced by the measurements. Including the effects of thermal dissipation reduces and eventually erases this memory effect.

The Wigner function (\ref{eq:WignerDef}) of the membrane after $n$ measurements can be derived from (\ref{eq:FinalDMatrix}) and is found to be
\begin{widetext}
\begin{eqnarray}
\label{eq:FinalWigFun}
W(x,p) &=&\frac{1}{{\mathrm tr}(\rho_{sys}(t))}\sum_{\gamma,\alpha_1,...,\beta_n}M^{(n)}_{\gamma\ \alpha_n}
M^{(n-1)}_{\alpha_n\alpha_{n-1}}...M^{(1)}_{\alpha_2\alpha_1}\rho_{\alpha_1\beta_1}
M^{(1)\dag}_{\beta_1\beta_2}...M^{(n-1)\dag}_{\beta_{n-1} \beta_n}
M^{(n)\dag}_{\beta_n \gamma}\nonumber\\
&\times&\exp\left(-i\sum_{i=1}^n t_i(\Omega_{L0}(\alpha_i-\beta_i)-\delta\Omega_{L}(\alpha_i^2-\beta_i^2))\right )\frac{1}{\pi\hbar}\tanh\eta\nonumber\\
&\times&\exp\left[-\frac{m\omega_m}{\hbar}\left\{\left(\left(x+\frac{A}{2}(X[\alpha]+X[\beta])\right)^2+\left(\frac{p}{m\omega_m}+\frac{A}{2}(P[\alpha]+P[\beta])\right)^2\right)\tanh\eta\right. \right. \nonumber\\
&+&\left. \left. iA(P[\alpha]-P[\beta])\left(x+\frac{A}{2}(X[\alpha]+X[\beta])\right)-iA(X[\alpha]-X[\beta])\frac{p}{m\omega_m}+i\frac{A^2}{2}(\phi[\alpha]-\phi[\beta])\rule{0pt}{16pt}\right\}\rule{0pt}{18pt}\right],
\end{eqnarray}
\end{widetext}
a form that clearly illustrates the shift in position and momentum of the various terms in the sum.

\subsection{Numerical results}

One simple way to detect the effect of quantum backaction on the state of the membrane is to consider a sequence of 2 measurements carried out in succession at times $t_1$ and $ t_2$, and to compare the outcome of the last measurement to the outcome of a measurement at that same time $t_2$, but skipping the first measurement  at $t_1$.  As a concrete example, we consider, for the parameters of the previous section, the following two scenarios:

1. $ F_y$ is measured once, at $t_2=\pi/\omega_m$ (i.e., half the oscillator period);

2. $F_y$ is measured first at $t_1=\pi/2\omega_m$ and then at $t_2=\pi/\omega_m$.

By simply calculating the trace of Eq.~(\ref{eq:FinalDMatrix}) for the 3 possible operators $ M^{(1)}$'s in scenario 1 we obtain the probabilities
\begin{eqnarray}
P(F_y=\pm 1)&=&0.375\nonumber\\
P(F_y=0)&=&0.25
\end{eqnarray}
In the second scenario 9 traces need to be evaluated instead of just 3. Summing then the three terms corresponding to the same final value of $ F_y$, so as to obtain the total probability of obtaining a particular value for the second measurement, we find
\begin{eqnarray}
P(F_y=\pm 1)&=&0.344\nonumber\\
P(F_y=0)&=&0.312
\end{eqnarray}

The simple example shows that backaction of the intermediate measurement should be readily observable, since it changes the probability for the three possible outcomes of the measurements at $t_2$ by a significant amount, of the order of several percent.

A slightly more complete look into the effects of intermediate measurements is provided by the Wigner functions shown in Figures~\ref{fig:measureFy1diffA} and \ref{fig:WigWigWig}. In Fig. ~\ref{fig:measureFy1diffA}, we show a series of plots for the case of a single measurement at $t=\pi/\omega_m$, for different values of $A$ ($0.01 A_0, 0.1 A_0,$ and $A_0$, respectively). In Fig.~\ref{fig:WigWigWig} we increase the number of measurements from 1 to 2. The time of the final measurement is the same as in Fig.~\ref{fig:measureFy1diffA}, $\pi/\omega_m$, but an intermediate measurement is made at time $t_1=\pi/2\omega_m$. The outcome of the final spin measurement is again $F_y = 1$, but, for the sake of direct comparison with Fig.~\ref{fig:measureFy1diffA}, we average over all possible results of the intermediate measurement. This is equivalent to an experiment in which the outcomes of the intermediate measurement are discarded or ignored. As expected, the Wigner functions where an intermediate measurement are noticeably different than the one lacking an intermediate measurement, hinting at a simple way to characterize the impact of quantum backaction on the membrane dynamics.

\section{Initial Coherent State}
We now turn to the situation where the center-of-mass state of the membrane is a coherent state, an initial condition that can be prepared by driving the membrane with a classical force. The main result of this section is to demonstrate that after $n$ measurements, the center-of-mass of membrane is split into a superposition of up to $3^n$ discrete coherent states, hinting at the possibility of generating a macrsocopic Schr{\"o}dinger cat.

\subsection{Evolution of the Coherent State}
The assumption that the membrane is initially in a coherent state allows us to eschew the density matrix formalism for the moment, and the initial state of the composite system is given by
\begin{eqnarray}
\label{eq:InitCoState}|\Psi(0)\rangle & =& D(a_0+ib_0) |0_{\rm mem}\rangle\otimes\sum_\alpha c_\alpha|\alpha\rangle
\end{eqnarray}
where  $\sum_\alpha |c_\alpha|^2=1$ and
\begin{equation}
D(u)=\exp(u\hat a^\dagger - u^*\hat a)
\end{equation}
is the displacement operator for the center-of-mass state of the membrane. Following a procedure similar to that for the initially thermal membrane, we arrive at the final state for the hybrid membrane-BEC system (see Appendix C)
\begin{widetext}
\begin{figure}[ht]
\begin{center}
$\begin{array}{ccc}
\includegraphics[width=2.1in]{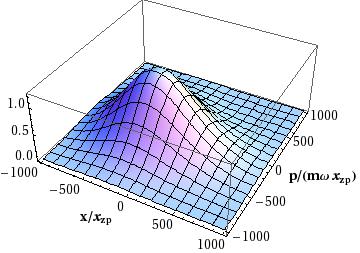}  &
\includegraphics[width=2.1in]{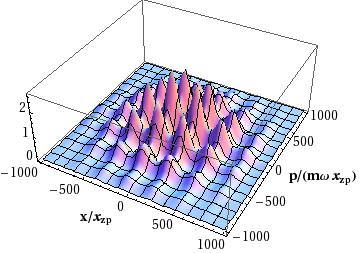} &
\includegraphics[width=2.1in]{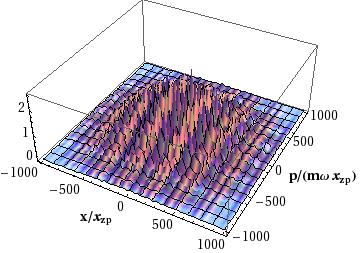} \\[0in]
\mbox{\bf a. $A= 0.01 A_0$} &\mbox{\bf b. $A= 0.1 A_0$} &\mbox{\bf c. $A= A_0$}\\[0.0cm]
\end{array}$
\end{center}
\caption{(Color online) Wigner distribution function of the membrane in the case of two measurements: the final spin measurement, at time $\pi/\omega_m$, yields the outcome  $F_y=1$ . As in Fig.~(\ref{fig:measureFy1diffA}), the three Wigner functions are plotted for increasing values of the backaction parameter, $A$, and the final Wigner functions are averaged over all possible outcomes of the intermediate measurement. }
\label{fig:WigWigWig}
\end{figure}
\begin{eqnarray}
\label{eq:FinalCoState}|\Psi(0)\rangle \hspace{-4pt}& =&\hspace{-14 pt}\sum_{\alpha,\alpha_1,...,\alpha_n}\hspace{-12 pt}M^{(n)}_{\alpha\ \alpha_n}
...M^{(1)}_{\alpha_2\alpha_1}c_{\alpha_1}\exp\hspace{-2pt}\left(\hspace{-2pt}-i\sum_{i=1}^nt_i\left(\Omega_{L0}\alpha_i\hspace{-2pt}-\delta\Omega_L\alpha_i^2+\frac{\omega_m}{2}\right)\hspace{-2pt}-i\frac{A\Theta[\alpha]}{2x_{zp}}\right)\hspace{-2pt} D(a[\alpha]+ib[\alpha])|0_{mem}\rangle\hspace{-1pt}\otimes\hspace{-1pt}|\alpha\rangle
\end{eqnarray}
with
\begin{eqnarray}
\label{eq:Definea}a[\sigma]&=&a_0\cos T_{1,n}+b_0\sin T_{1,n}-\frac{A}{2x_{zp}}X[\sigma],\\
\label{eq:Defineb}b[\sigma]&=&b_0\cos T_{1,n}-a_0\sin T_{1,n}-\frac{A}{2x_{zp}}P[\sigma],\\
\label{eq:DefineTheta}\Theta[\sigma]&=&\sum_{i=1}^n\sigma_i \left[ a_0(\sin T_{1,i}-\sin T_{1,i-1})
-b_0(\cos T_{1,i}-\cos T_{1,i-1})
+\frac{A}{2x_{zp}}\sum_{j=1}^i(\sigma_j-\sigma_{j-1})(\sin T_{j,i}-\sin T_{j,i-1}) \right ],
\end{eqnarray}
\end{widetext}
and all other definitions are as before. This wave-function is not normalized; the probability of obtaining a particular sequence of measurement outcomes is given by $\langle\Psi(t)|\Psi(t)\rangle$.

Because the real part of the displacement is proportional to $\langle x\rangle$ and the imaginary part is proportional to $\langle p\rangle$, we have a very clear picture of the physics arising in this situation. The membrane's initial displacement oscillates back and forth semi-classically. However, each time the BEC's spin is measured, the membrane's wave function splits into three distinct components, each having received a different kick from its interaction with the different possible spin orientations of the BEC. This would indicate that the membrane was indeed put into a cat state, except for the fact that the kick is proportional to $A$, which is typically at most comparable in size to the zero point oscillations $x_{zp}$. If, going forward, we can increase $A$ in an experimentally realizable setting, this system may provide an excellent demonstration of a macroscopic object put into a highly non-classical state.

\subsection{Wigner function}

Following a sequence of measurements, the Wigner function of the initially coherent state of the membrane motion becomes
\begin{widetext}
\begin{eqnarray}
\label{eq:TrulyFinalWigFun}
W(x,p) &=&\frac{1}{\pi\hbar\langle\Psi(t)|\Psi(t)\rangle}\sum_{\gamma,\alpha_1,...,\beta_n}M^{(n)}_{\gamma\ \alpha_n}
M^{(n-1)}_{\alpha_n\alpha_{n-1}}...M^{(1)}_{\alpha_2\alpha_1}\rho_{\alpha_1\beta_1}
M^{(1)\dag}_{\beta_1\beta_2}...M^{(n-1)\dag}_{\beta_{n-1} \beta_n}
M^{(n)\dag}_{\beta_n \gamma}\nonumber\\
&\times&\exp\left(-i\sum_{i=1}^n t_i(\Omega_{L0}(\alpha_i-\beta_i)-\delta\Omega_{L}(\alpha_i^2-\beta_i^2))\right )\nonumber\\
&\times&\exp\left[-\frac{m\omega_m}{\hbar}\left\{\left(x-x_{zp}(a[\alpha]+a[\beta]+i(b[\alpha]-b[\beta]))\right)^2+\left(\frac{p}{m\omega_m}-x_{zp}(b[\alpha]+b[\beta]-i(a[\alpha]-a[\beta]))\right)^2 \right\}\right. \nonumber\\
&+& \left. i(a[\beta]b[\alpha]-a[\alpha]b[\beta])-i\frac{A}{2x_{zp}}(\Theta[\alpha]-\Theta[\beta])-\frac{1}{2}(a[\alpha]-a[\beta])^2-\frac{1}{2}(b[\alpha]-b[\beta])^2\rule{0pt}{18pt}\right]
\end{eqnarray}
\end{widetext}

Figure~\ref{fig:MorecoherentWigs} shows the post-measurement Wigner function after successive measurements. The key point here is that repeated measurements can lead to very non-classical states, as is evident from the resulting negative valued Wigner functions. As can be seen in Eqs.~(\ref{eq:FinalCoState}) and (\ref{eq:TrulyFinalWigFun}), each measurement introduces different phase factors to the initial coherent state, along with splitting it into different coherent states. It is the quantum interference between the different coherent states thus generated that gives rise to the non-classical Wigner functions. Relatively few measurements are required to generate such non-classical states. In fact, it is possible to turn a coherent state into a non-classical state after only one measurement. Fig.~\ref{fig:MorecoherentWigs} shows one such case. Here, $a_0=b_0=1$ and the result of the first measurement after a time interval of $t_1=\pi/\omega_m$ is $F_y=0$. Repeated measurements after equal time intervals lead to states that resemble displaced Fock states and other more complex states.

If the initial displacement is small in magnitude (as in Fig.~\ref{fig:MorecoherentWigs}), production of this non-classical behavior is rather insensitive to the measurement timing. For this particular case, with the outcome $F_y=0$, it is seen for $t_1$ in the approximate range $0.6 \pi/\omega_m - 1.8 \pi/\omega_m$. On the other hand if the initial displacement is large (for
example, $a_0=50\sqrt{2},\ b_0=0$), the non-positive Wigner function is much harder to find; in this case it only occurs from $t_1 \approx 0.77 \pi/\omega_m - 0.79 \pi/\omega_m$. Nevertheless, in either case, successive measurements can build up some very interesting non-classical states. It may be possible to obtain a high degree of probabilistic quantum state control of the membrane if the dissipative effects do not wash out the interference too quickly.

\section{Conclusion}
In conclusion, we have demonstrated that by coupling a magnetic membrane to a spinor BEC, we can monitor and manipulate the position of the membrane. A projective measurement procedure induces significant backaction that can be measured for reasonable experimental parameters. We investigated the effect of this interaction for different initial membrane states, namely thermal and coherent states. We discussed the possibility to measure backaction of a quantum measurement on the membrane via repeated measurements and the potential to generate cat states of the oscillator via such repeated measurements in the case of an initial coherent state. This would be a major accomplishment as such a state has not yet been prepared for a solid, macroscopic object.

As mentioned above, we will in future work look at the effects of using a more dispersive (and more experimentally plausible) measurement scheme based on observation of the Larmor precession. Additionally, we will study the effects of including dissipation on the system's dynamics. Other possibilities for extending the model even further include inclusion of a general coupling to other spin components, i.e. an interaction Hamiltonian of the form $\bf{F}\cdot\bf{B}$, because at short distances the $x$ and $y$ gradients of the magnetic field become significant. Also, the BEC is a spatially finite system, so we may wish to exploit the ability to measure multiple ``pixels'' of the condensate in order to gain better information about the membrane's position and/or better control of the membrane's state.

\begin{acknowledgments}
We thank David Brown for his help during the initial stages of these calculations. This work was supported by the DARPA QuASAR program through a grant from AFOSR and the DARPA ORCHID program through a grant from ARO, the US Army Research Office, and by NSF. M. V. acknowledges support from the Alfred P. Sloan Foundation. MET is supported by TUBITAK, The Scientific and Technological Research Council of Turkey. 
\end{acknowledgments}
\appendix
\section{Derivation of interaction Hamiltonian}
Our starting assumption is that the magnetic domain on the membrane is a point dipole located at the origin. 
\begin{widetext}
\begin{figure}[t!!]
\begin{center}
$\begin{array}{cc}
\includegraphics[width=2.75in]{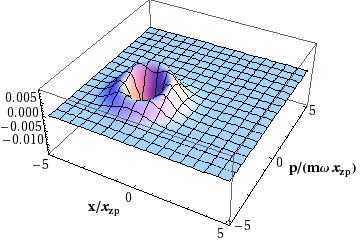}  &
\includegraphics[width=2.75in]{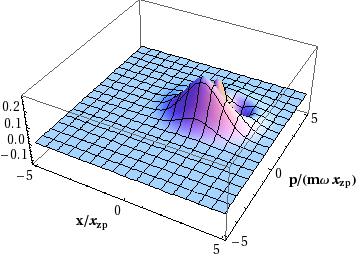}\\[0in]
\mbox{\bf a. 1st meas. $F_y=0, t_1=\pi/\omega_m$ } &\mbox{\bf b. 2nd meas. $ F_y=1, t_2=\pi/\omega_m$}\\[0in]
\includegraphics[width=2.75in]{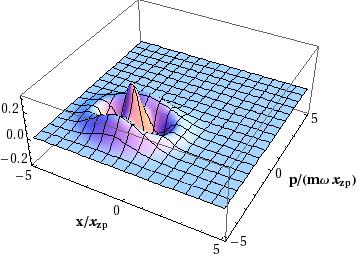} &
\includegraphics[width=2.75in]{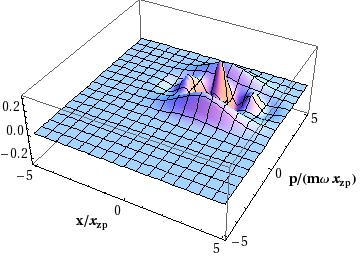}\\[0in]
\mbox{\bf c. 3rd meas. $ F_y=-1, t_3=\pi/\omega_m$} &
\mbox{\bf d. 4th meas. $ F_y=1, t_4=\pi/\omega_m$}\\[0.0cm]
\end{array}$
\end{center}
\caption{(Color online) Post-measurement Wigner distribution functions of the membrane, initially in a coherent state $\alpha=1+i1$. We notice that repeated measurements do indead lead to some interesting non-classical states. The measurement results and evolution time before measurement are given for each plot.}
\label{fig:MorecoherentWigs}
\end{figure}
\end{widetext}
The magnetic field due to that dipole is
\begin{equation}
{\bf B(r)}=\frac{\mu_0}{4 \pi}\frac{1}{r^3}\left(3({\bf \mu_m.\hat{r}}){\bf\hat{r}}-{\bf \mu_m}\right)
\label{eq:Bfield}
\end{equation}
It is the small inhomogeneity of the resulting magnetic field along the long axis $z$ of the condensate, and the resulting variation in Larmor precession frequency, that permits us to characterize the center-of-mass mode of oscillation of the membrane. 

For a dipole polarized along the $z$-axis (${\bf \mu_m} = \mu_m{\bf\hat{z}}$),  at distance $r=(x^2+y^2+z^2)^{1/2}$, the components of the magnetic field (\ref{eq:Bfield}) are
\begin{eqnarray}
B_x(x,y,z)&=&\frac{\mu_0}{4\pi}\frac{\mu_m}{r^{3}}\left[\frac{3xz}{r^2}\right],\\
B_y(x,y,z)&=&\frac{\mu_0}{4\pi}\frac{\mu_m}{r^{3}}\left[\frac{3yz}{r^2}\right],\\
B_z(x,y,z)&=&\frac{\mu_0}{4\pi}\frac{\mu_m}{r^{3}}\left[\frac{3z^2}{r^2}-1\right].
\end{eqnarray}
Let $x=x_0+x_m$, where $x_0$ is the equilibrium value of $x$ for a condensate atom, $x=0$ is the equilibrium position of the membrane, and $x_m$ is the small sinusoidal displacement of the membrane around the origin. Expanding the expression for magnetic fields for small $x_m$ (up to first order) we get
\begin{eqnarray}
B_x &\approx& \frac{\mu_0\mu_m}{4\pi r_0^{5}}\left[ 3x_0z-\frac{3z(4x_0^2-y^2-z^2)}{r_0^2}x_m \right],\\
B_y &\approx& \frac{\mu_0\mu_m}{4\pi r_0^{5}}\left[ 3yz- \frac{15x_0yz}{r_0^2}x_m \right],\\
B_z &\approx& \frac{\mu_0\mu_m}{4\pi r_0^{5}}\left[ (2z^2-x_0^2-y^2)\right.\\ \nonumber
&&\left.+\frac{3x_0(x_0^2+y^2-4z^2)}{r_0^2}x_m \right].
\end{eqnarray}
Here, $r_0=(x_0^2+y^2+z^2)^{1/2}$. Assuming a two-dimensional condensate (i.e. $y\approx0$), we can set $B_y\rightarrow 0$. For theoretical simplicity, we assume that the measurements are made on the part of the condensate directly above the magnet, i.e. close to $z=0$. Under these approximations, the magnetic field simplifies to being predominantly in the z-direction,
\begin{equation}
 B_z\approx\frac{\mu_0\mu_m}{4\pi x_0^{4}}\left[-x_0 + 3x_m \right].
\end{equation}
Combining this with the quantization field $B_0$, we end up with the interaction Hamiltonian of the form
\begin{equation}
V=\mu_B g_F F_z \left[B_0-\frac{\mu_0\mu_m}{4 \pi x_0^3}+\frac{3\mu_0\mu_m}{4 \pi x_0^4}x_m\right].
\end{equation}
We can break this total magnetic field into a constant part, $B_c$ (sum of the first two terms), and a component that varies at the membrane frequency, $B_v'x_m$. Here, the gradient of the magnetic field is
\begin{equation}
 B_v'=\frac{3\mu_0\mu_m}{4 \pi x_0^4}.
\end{equation}
This is the form used in Eq.~(\ref{eq:simpleAHamiltonian}).

\begin{widetext}
\section{Derivation of Successive Measurement Density Matrix}
Here we outline the steps used in section IV to derive Eq.~(\ref{eq:FinalDMatrix}). We arrive at the result via induction. Suppose that $n-1$ measurements have been performed already. We denote the density matrix at this moment by $\rho^{(n-1)}$. After the next period of free evolution and the $n$th measurement, we have
\begin{eqnarray}
\label{eq:RecurSetup}
\rho^{(n)} &=& M^{(n)}U(t_{n})\rho^{(n-1)}U^\dag(t_{n})M^{(n)\dag}.
\end{eqnarray}
To find the matrix elements, we insert completeness relationships
\begin{eqnarray}
\label{eq:RecurComRel}
\langle \alpha,x_f |\rho^{(n)}|\beta,x_i\rangle &=& \sum_{\alpha_n}\sum_{\beta_n}\int dx\int dy\langle \alpha,x_f |M^{(n)}U(t_n)|\alpha_n,x\rangle\langle\alpha_n,x|\rho^{(n-1)}|\beta_n,y\rangle\langle\beta_n,y|U^\dag(t_n)M^{(n)\dag}|\beta,x_f\rangle.
\end{eqnarray}
Now, the propagator can be derived from Eq.~(\ref{eq:thermalHO}) by multiplying by the trace, substituting $\eta \rightarrow i \omega_mt/2$, and including the contributions of the $F_z$-dependent parts of the Hamiltonian. Using these techniques, we arrive at the needed matrix elements:
\begin{eqnarray}
\label{eq:ModifiedProp}
\langle \sigma,u |M^{(n)}U(t_n)|\tau,v\rangle\ &=& M^{(n)}_{\sigma\ \tau}\sqrt\frac{8 m \omega_m}{i\pi\hbar\sin\omega_m t}\exp\left [-i \left[t_n\Omega_{L0}\tau-t_n\delta\Omega\tau^2\rule{0pt}{14pt}\right.\right.\nonumber\\
&+&\frac{m\omega}{4\hbar}\left.\left.\left((u+v+2A\tau)^2\tan\frac{\omega_m t_n}{2}-(u-v)^2\cot\frac{\omega_m t_n}{2}\right)\right]\right ].
\end{eqnarray}
The fact that Eq.~(\ref{eq:thermalHO}) and the individual terms of Eq.~(\ref{eq:condHOdensity}) (i.e. terms of the form of Eq.~(\ref{IntSysDMatrix})) do not differ except for shifts in the coordinates and some complex phase shifts leads us to conjecture that additional evolution and measurements will not alter the underlying structure of the density matrix. Thus, we try a matrix of the form (\ref{eq:FinalDMatrix}) for $\rho^{(n-1)}$, but with unknowns for $X^{(n-1)}$, etc. (superscript added for clarity) and see if evaluation of the integral (\ref{eq:RecurComRel}) produces a similar output for $\rho^{(n)}$. Indeed it does, and it also gives us the recurrence relations used to derive Eqs.(\ref{eq:DefineX}-\ref{eq:DefinePhi}). These are
\begin{eqnarray}
\label{eq:RecurRel}
X^{(n)}[\sigma]&=&X^{(n-1)}[\sigma]\cos\omega_m t_n + P^{(n-1)}[\sigma]\sin\omega_m t_n + \sigma_n(1-\cos\omega_m t_n),\\
P^{(n)}[\sigma]&=&P^{(n-1)}[\sigma]\cos\omega_m t_n - X^{(n-1)}[\sigma]\sin\omega_m t_n + \sigma_n\sin\omega_m t_n,\\
\phi^{(n)}[\sigma]&=&\phi^{(n-1)}[\sigma]+(X^{(n)}[\sigma]^2+X^{(n-1)}[\sigma]^2)\cot\omega_m t_n-2X^{(n)}[\sigma]X^{(n-1)}[\sigma]\csc\omega_m t_n.
\end{eqnarray}
The solutions to these recurrence relationships are the equations given above.

\section{Derivation of Coherent State Evolution}
We again will arrive at the result (\ref{eq:FinalCoState}) by using induction and finding a set of recurrence relations as in Appendix B, but the steps involved are slightly different. We first split the Hamiltonian (\ref{eq:Hamiltonian}) into two parts $\tilde H_{\rm BEC}$ and $(\tilde H_m + \tilde V)$. These commute, so the unitary evolution operator can simply be factorized into two operators, one acting only on the BEC and another acting as a shifted harmonic oscillator. We also note that the unperturbed membrane Hamiltonian and $\tilde H_m + \tilde V$ are related by a simple unitary transformation, namely
\begin{eqnarray}
\tilde H_m + \tilde V &=& D\left(-\frac{A}{2x_{zp}}F_z\right)H_mD\left(\frac{A}{2x_{zp}}F_z\right).
\end{eqnarray}
Using this fact and the well-known results
\begin{eqnarray}
\exp[-iH_mt/\hbar]D(\zeta)&=&D(e^{-i\omega_mt}\zeta)\exp[-iH_mt/\hbar],\\
D(\zeta)D(\xi)&=&D(\zeta+\xi)e^{(\zeta\xi^*-\zeta^*\xi)/2},
\end{eqnarray}
we can postulate that the system after $n-1$ measurements is specified by a superposition of states as in (\ref{eq:FinalCoState}), apply an additional evolution and measurement,
\begin{eqnarray}
|\Psi^{(n)}\rangle&=& M^{(n)}\tilde U_{\rm BEC}(t_n)\tilde U_m(t_n)|\Psi^{(n-1)}\rangle,
\end{eqnarray}
and then use the result to find the following recurrences:
\begin{eqnarray}
a^{(n)}[\sigma]&=&a^{(n-1)}[\sigma]\cos\omega_m t_n + b^{(n-1)}[\sigma]\sin\omega_m t_n - \frac{\sigma_nA}{2x_{zp}}(1-\cos\omega_m t_n),\\
b^{(n)}[\sigma]&=&b^{(n-1)}[\sigma]\cos\omega_m t_n - a^{(n-1)}[\sigma]\sin\omega_m t_n - \frac{\sigma_nA}{2x_{zp}}\sin\omega_m t_n,\\
\Theta^{(n)}&=&\Theta^{(n-1)}+\alpha_n(b^{(n-1)}-b^{(n)}).
\end{eqnarray}
Note the strong resemblance between the first two of these equations and those from the previous appendix. This is to be expected, as the real and imaginary parts of the displacement are proportional to the expected position and momentum of the coherent state, respectively. The solutions to these equations are Eqs.~(\ref{eq:Definea}-\ref{eq:DefineTheta}).

\end{widetext}

\section{Experimental Implementation}

Here, we provide details on the experimental detection of the quantum backaction on the micromechanical membrane. 
The membrane is composed of Silicon Nitride with a micron-scale magnetic domain deposited at its center. These structures are fabricated by coating silicon wafers with around 100 nm of high stress silicon nitride and then etching the silicon in a 100 $\mu$m $\times$ 100 $\mu$m region to reveal a free standing membrane. Quality factors exceeding $10^6$ have been demonstrated with similar SiN membranes for resonance frequencies in the range of 500 kHz~\cite{mukundref2}.

The magnetic domain is deposited on this membrane by a combination of photolithography and sputtering of a high permeability material like permalloy~\cite{mukundref1}. Peak surface fields on the order of 1 T at the surface of this magnetic domain should ensure field gradients around $10^3$ T/m in the regions above the domain. The membrane is supported on a cryogenically cooled flange housed in a UHV chamber. Sputtered gold films deposited on the substrate in the regions surrounding the membrane serve as mirrors for trapping and imaging. Spinor Bose condensates of $^{87}$Rb in the $F=1$ hyperfine state will be optically confined about 5 $\mu$m above the membrane in a quasi-2D trap created by shallow angle interference of laser beams reflected off the gold surface~\cite{mukundref3}. This ensures a quasi-2D confinement of the condensates such that the spatial extent of the gas in the direction normal to the membrane is less than the spin healing length $\xi = \sqrt{\hbar/2 m |c_2 n|}$. Here, $c_2 = 4 \pi\hbar^2 (a_2 - a_0)/m$ is the spin dependent coupling strength of the spinor gas and $a_0 (a_2)$ is the $s$-wave scattering length in the $F=0 (F=2)$ channel. In this way, we can freeze out spin dynamics along this dimension~\cite{Vengalatorre2007}.

The estimation of the membrane's position is based on detecting the Zeeman shift at the location of the condensate due to the micromotion of the membrane. This is achieved by magnetization-sensitive phase contrast imaging as demonstrated in~\cite{Vengalatorre2007}. To summarize this technique briefly, a sequence of non-destructive phase contrast images of the condensate are obtained with far off-resonant circular polarized light. Due to Larmor precession of the gas, the phase imprinted on the probe light is modulated at the Larmor frequency leading to a phase contrast signal given by
\begin{equation}
\label{eq:pcisig}
s = 1 + 2 \tilde{n} \sigma_0 (\gamma/2 \Delta) [a_0 + a_1 \langle F_y \rangle + a_2 \langle F_y^2 \rangle]
\end{equation}
where $\tilde{n}$ is the column density of the gas, $\sigma_0 = 3 \lambda^2/2 \pi$ is the resonant cross section, $\Delta$ is the detuning of the probe light from resonance and $\gamma$ is the natural linewidth. $F_y$ is the projection of the local atomic spin along the imaging axis. The constants $a_0, a_1$ and $a_2$ depend on the detuning of the probe light and describe the isotropic polarization and optical activity of the condensate. By analyzing the sequence of phase contrast images on a pixel-by-pixel basis, we can estimate the local Larmor precession rate and hence, the local magnetic field. At the shot noise limit, we estimate a magnetic field sensitivity on the order of 15 fT/Hz$^{1/2}$ for a condensate of area 50 $\mu$m$^2$ in the plane of the membrane~\cite{mukundref5}. For the expected field gradient of $10^3$ T/m at the location of the condensate, this field sensitivity translates to a position sensitivity of $15 \times 10^{-18}$ m/Hz$^{1/2}$. This is much smaller than the zero point motion of the oscillator which has an amplitude on the order of $6 \times 10^{-15}$ m. 



\begin{thebibliography}{10}

\bibitem{Braginsky}
V. B. Braginsky and F Ya. Khalili, ``Quantum Measurement'', Cambridge University Press (1992).

\bibitem{KippenbergScience09}
T.J. Kippenberg and K.J. Vahala, Science {\bf 321}, 1172 (2008).
F. Marquardt and S. M. Girvin, Physics {\bf 2}, 40 (2009).

\bibitem{Aspect82}
Alain Aspect, Jean Dalibard, and G\'{e}rard Roger, Phys. Rev. Lett. {\bf 49}, 1804 (1982).

\bibitem{Hollenhorst}
James N. Hollenhorst , Phys. Rev. D {\bf 19}, 1669 (1979).

\bibitem{Kitching2006}
Ying-Ju Wang \textit{et. al}, Phys. Rev. Lett. {\bf 97}, 227602 (2006).

\bibitem{Treutlein2010}
D. Hunger \textit{et. al}, Phys. Rev. Lett. {\bf 104}, 143002 (2010).

\bibitem{Treutlein2007}
P. Treutlein  \textit{et. al}, Phys. Rev. Lett. {\bf 99},  140403  (2007).

\bibitem{Genes2008}
C. Genes, D. Vitali and P. Tombesi, Phys. Rev. A {\bf 77}, 050307(R) (2008).

\bibitem{Hammerer2009}
K. Hammerer \textit{et. al}, Phys. Rev. Lett. {\bf 103}, 063005 (2009).

\bibitem{SinghPM2010}
S. Singh and P. Meystre, Phys. Rev. A {\bf 81}, 041804(R) (2010).

\bibitem{Hammerer2010}
K. Hammerer \textit{et. al},Phys. Rev. A, {\bf 82}, 021803(R) (2010).

\bibitem{Vengalatorre2007}
M. Vengalatorre \textit{et. al}, Phys. Rev. Lett. {\bf 98}, 200801 (2007).

\bibitem{Obrecht2007}
J. M. Obrecht \textit{et. al}, Phys. Rev. Lett. {\bf 98}, 063201 (2007).

\bibitem{Zoest2010}
T. van Zoest \textit{et. al}, Science {\bf 328}, 1540 (2010)

\bibitem{Higbie2005}
J. M. Higbie \textit{et. al}, Phys. Rev. Lett. {\bf 95}, 050401 (2005).

\bibitem{Murch2008}
K. W. Murch {\it et. al}, Nature Physics {\bf 4}, 561 (2008) 

\bibitem{mukundref1}
M. Vengalattore \textit{et al}, J. Appl. Phys. {\bf 95}, 4404 (2004).

\bibitem{mukundref2}
B. M. Zwickl  \textit{et al}, Appl. Phys. Lett. {\bf 92}, 103125 (2008). 

\bibitem{mukundref3}
D. Gallego \textit{et al}, Opt. Lett. {\bf 34}, 3463 (2009).

\bibitem{mukundref5}
M. Vengalattore \textit{et al}, (to be published). 

\end{thebibliography}
\end{document}